\documentstyle[multicol,aps,prb,epsf]{revtex}

\begin{document}
\pagestyle{empty}

\newcommand{\bc}{\begin{center}}
\newcommand{\ec}{\end{center}}
\newcommand{\be}{\begin{equation}}
\newcommand{\ee}{\end{equation}}
\newcommand{\beqn}{\begin{eqnarray}}
\newcommand{\eeqn}{\end{eqnarray}}

\begin{multicols}{2}
\narrowtext
\parskip=0cm

\noindent
{\large\bf Comment on "Statistical Mechanics of Membrane-Protein Conformation:
A Homopolymer Model"}
\smallskip

In a recent Letter Park and Sung \cite{ps} modeled a membrane-protein 
conformation by grand canonical ensemble approach using a homopolymer model.
In this comment I 
would like to point out that although their model has no serious physical
mistake, it is far from biological reality.

The basic scenario of their model is depicted in Fig.1(a). 
Therefore, they expand the interaction into loops, inclusions and adsorptions.
It is unrealistic in four aspects: Firstly, for a real membrane
the chemical potential inside and outside the cell should be different.
However, this point is not included in their fugacities.
Secondly, although real protein sequence 
is head-tail distinguishable and the authors
tried to distinguish it in their calculation (Eq. (6)),
however, their mathematical model 
does not include this effect because $Q_S$, $Q_L$ and $Q_H$
are only classical numbers which are commutable.
Thirdly, They consider  adsorptions and inclusions to be of the same order
processes. However, it is questionable
because the number of hydrogen bonds involved in an inclusion can vary.
Furthermore, the last diagram in their Fig.2 is unlikely to happen because
it requires much too high total free energy, i.e. include entropy effects,
to activate. 
Fourthly, and most importantly, their model is inconsistent with 
a three-dimensional structural finding, to a resolution of $1.9 \AA$,
of the {$\alpha$}-hemolysin pore almost two years ago\cite{x}.

The first point is allright as a first order approximation of the cellular 
environment.
However, the second point is mathematically wrong.
On the other hand, the structural
finding told us that protein should looks like sperms with a big 'head' which
consists of about $80\%$ of the total animo acids.
Therefore,
the head-tail distinguishibiltiy is important.
It can be considered as
a collective partition function different from the partition
function of the other part of the protein chain.
Furthermore, the 'tail' (in the authors' terminology)
is unlikely to form a second adsorption on the membrane
surface. It should be folded back to itself to form a 'head' instead, if
one thinks in terms of linear polymers.

The structural finding told us that protein aggregates to form a pore due to
several protein's tail penetrating the membrane (Fig.1b) instead of one protein
chain winding back and forth on the membrane (Fig.1a).
These protein units are bound together
because of extensive hydrophobic and hydrophilic contacts of their heads.
Specifically, for the sample studied in Ref.2, it is seven pieces gathered
together to form a mushroom-like pore.
The reader
should consult Ref.2 for better illustrations and thorough explanation.

In conclusion it turns out nature operates much more efficiently than the 
authors expected. 
The structural finding told us that protein penetrating the membrane only 
as a 'zeroth order' process in the authors'
formulation, i.e., only single inclusion and
single adsorptions are possible. 
Furthermore, these two processes actually happen in sequence. They are
not separate processes. Therefore there will not have the problem about
which one have higher energy.

\begin{figure}
\epsfxsize=\columnwidth\epsfbox{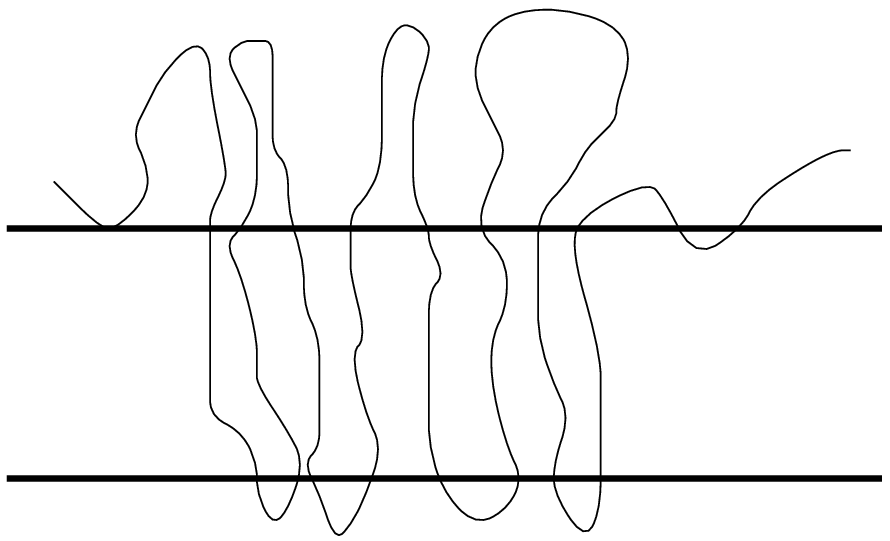} 
\vspace{0.5cm}
\epsfxsize=\columnwidth\epsfbox{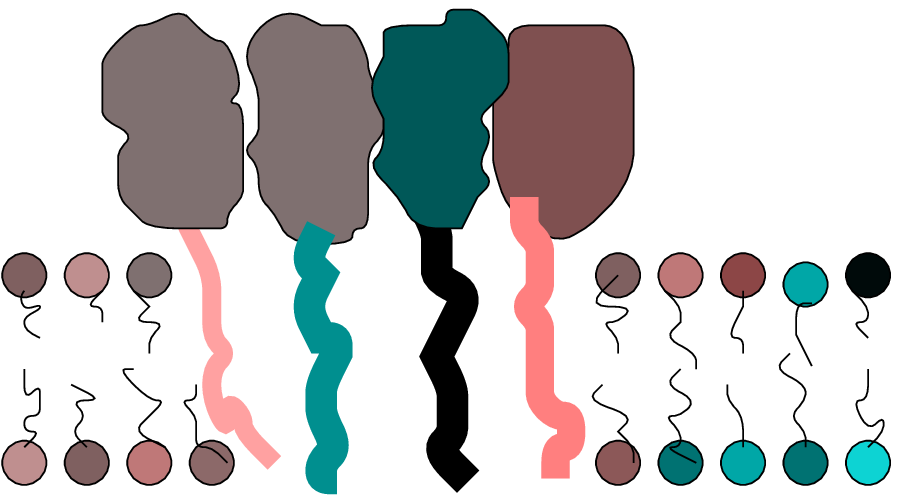}
\vspace{0.3cm}
\caption{(a) People though membrane-protein interacts like chop-sticks and 
spaghetti. But how could such high order interactions happen in nature ? 
(b) With better resolution, structural finding has resolved definitely that
the interaction is only 'zeroth order', but with many protein units. 
\label{fig}
}
\end{figure}

I thank P. J. Park and W. Sung for
the preprint of their work.
This work is partially supported by APEC fellowship.

\bigskip
\noindent
{Julian Juhi-Lian Ting}

{\small
Department of Physics,

Pohang University of Science and Technology,

Pohang, 790-784 Korea
}

\bigskip
\noindent
Date: \today


\end{multicols}

\end{document}